\documentclass{PoS}

\newcommand{\be}{\begin{equation}}
\newcommand{\ee}{\end{equation}}
\newcommand{\bea}{\begin{eqnarray}}
\newcommand{\eea}{\end{eqnarray}}
\newcommand{\bi}{\begin{itemize}}
\newcommand{\ei}{\end{itemize}}
\newcommand{\bc}{\begin{center}}
\newcommand{\ec}{\end{center}}

\title{Identification of shallow two-body bound states in finite volume}

\ShortTitle{Identification of shallow two-body bound states in finite volume}

\author{\speaker{Shoichi Sasaki}%
         \thanks{The results of calculations were performed by using of RIKEN Super Combined Cluster (RSCC). S.S. is supported by the JSPS for a Grant-in-Aid for Scientific Research (C) (No. 19540265).}\\
        Department of Physics, University of Tokyo, Hongo, Tokyo 113-0033, JAPAN\\
        E-mail: \email{ssasaki@phys.s.u-tokyo.ac.jp}}

\author{Takeshi Yamazaki\\
        Physics Department, University of Connecticut, Storrs, CT 06269-3046, USA\\
        E-mail: \email{yamazaki@phys.uconn.edu}}

\abstract{
\indent\indent
We discuss signatures of bound-state formation
in finite volume via the L\" uscher finite size method.
Assuming that the phase-shift formula in this method inherits all aspects of
the quantum scattering theory, we may expect that the bound-state formation 
induces the sign of the scattering length to be changed.
If it were true, this fact provides us a distinctive identification of a shallow 
bound state even in finite volume through determination of whether the second
lowest energy state appears just above the threshold.
We also consider the bound-state pole condition
in finite volume, based on L\" uscher's phase-shift formula 
and then find that the condition is fulfilled only in the infinite volume limit, 
but its modification by finite size corrections
is exponentially suppressed by the spatial lattice size $L$. 
These theoretical considerations are also numerically checked  
through lattice simulations to calculate the positronium spectrum 
in compact scalar QED, where the short-range interaction between 
an electron and a positron is realized in the Higgs phase.
}

\FullConference{The XXV International Symposium on Lattice Field Theory\\
		 July 30 - August 4 2007\\
		 Regensburg, Germany}

\begin{document}

\section{Introduction}

Signatures of bound-state formation {\it in finite volume} are
of main interest in this paper. In the infinite volume, the bound state is well 
defined since there is no continuum state below threshold. However, in a finite box 
on the lattice, all states have discrete energies. Even worse, the lowest energy level 
of the elastic scattering state appears below threshold in the case if an interaction 
is attractive between two particles~\cite{Luscher:1985dn}. 
Therefore, there is an ambiguity to distinguish between the shallow (near-threshold) 
bound state and the lowest scattering state in finite volume in this sense.

We may begin with a naive question: what is the legitimate 
definition of the shallow bound state in the quantum mechanics? 
In the scattering theory, 
poles of the $S$-matrix or the scattering amplitude
correspond to bound states~\cite{Newton:1982qc}. 
It is also known that the appearance of the $S$-wave 
bound state is accompanied by an abrupt sign change of the $S$-wave scattering length~\cite{Newton:1982qc}.
It is interpreted that formation of one bound-state raises the phase shift 
at threshold by $\pi$. This particular feature is generalized as Levinson's 
theorem~\cite{Newton:1982qc}.
Thus, it is interesting to consider how the formation condition
of bound states is implemented in L\" uscher's finite size method, 
which is proposed as a general method for computing low-energy scattering 
phases of two particles in finite volume~\cite{Luscher:1985dn}. 

In this paper, we discuss bound-state formation on the basis of the L\" uscher's
phase-shift formula 
and then present our proposal for identifying the shallow bound state in finite volume.
To exhibit the validity and efficiency of our proposal, we perform numerical studies
of the positronium spectroscopy in compact scalar QED model.
In the Higgs phase of $U(1)$ gauge dynamics, the photon is massive. Then, 
massive photons give rise to the short-ranged interparticle force between 
an electron and a positron, which is exponentially damped. 
In this model, we can control positronium formation 
in variation with the strength of the interparticle force and then explore distinctive
signatures of the bound-state formation in finite volume. 
The contents of this paper are based on our published work~\cite{Sasaki:2006jn}.

\section{Bound-state formation in L\"uscher's formula}

In quantum scattering theory, the formation condition of bound states is implemented 
as a pole in the $S$-matrix or scattering amplitude. Here, an important question 
naturally arises as to how bound-state formation is studied through 
L\"uscher's phase-shift formula~\cite{Luscher:1985dn}.
Intuitively, the pole condition of the $S$-matrix: $S={\rm e}^{2i\delta_0(p)} = \frac{\cot\delta_0(p)+i}{\cot\delta_0(p)-i}$ is expressed as
\be
\cot \delta_0(p) = i,
\ee
which is satisfied at $p^2=-\gamma^2$ where positive real $\gamma$ represents the binding momentum. In fact, as we will discuss in the following, such a condition is fulfilled 
only in the infinite volume. However the finite-volume corrections on this pole condition 
are exponentially suppressed by the size of spatial extent $L$.

It was shown by L\"uscher that 
the $S$-wave phase shift $\delta_{0}$ can be calculated by measuring the relative 
momentum of two particles $p$ in a finite box $L^3$ with a spatial size $L$ through the relation
%
%
\be
\tan \delta_{0}(p)=\frac{\pi^{3/2} \sqrt{q^2}}{{\cal Z}_{00}(1,q^2)}\;\;\;\;{\rm at}\;\;
q=Lp_n/2\pi , 
\label{Eq.LucsherFormula}
\ee
where the generalized zeta function ${\cal Z}_{00}(s,q^2) \equiv \frac{1}{\sqrt{4\pi}}\sum_{{\bf n}\in Z^3}({\bf n}^2 - q^2)^{-s}$ is defined through
analytic continuation in $s$ from the region $s>3/2$ to $s=1$~\cite{Luscher:1985dn}.
For {\it negative} $q^2$, an exponentially convergent expression
of the zeta function ${\cal Z}_{00}(s, q^2)$ has been derived 
in Ref.~\cite{Elizalde:1997jv}.
For $s=1$, it is given by
\be
{\cal Z}_{00}(1,q^2)=-\pi^{3/2}\sqrt{-q^2}+\sum_{{\bf n}\in{\bf Z^3}}{}^{\prime}
\frac{\pi^{1/2}}{2\sqrt{\bf n^2}}e^{-2\pi \sqrt{-q^2{\bf n^2}}},
\label{Eq:ECSformula}
\ee
where $\sum^{\prime}_{{\bf n}\in Z^3}$ means the summation without ${\bf n}=(0,0,0)$.
We now insert Eq.~(\ref{Eq:ECSformula}) into Eq.~(\ref{Eq.LucsherFormula}) and then obtain
the following formula, which is mathematically equivalent to Eq.~(\ref{Eq.LucsherFormula})
for {\it negative} $q^2$: 
%
%
\be
\cot \delta_0(p) = i + \frac{1}{2\pi i}\sum_{{\bf n}\in{Z^3}}{}^{\prime}
\frac{1}{\sqrt{-q^2{\bf n}^2}}e^{-2\pi\sqrt{-q^2{\bf n}^2}}.
\label{Eq:BScond}
\ee
The second term in the r.h.s. of Eq.~(\ref{Eq:BScond}) vanishes in the limit 
of $q^2\rightarrow -\infty$. It clearly indicates that negative infinite 
$q^2$ is responsible for the bound-state formation. 
Therefore, in this limit, the relative momentum squared $p^2$ 
approaches $-\gamma^2$, which must be non-zero. 
Meanwhile, the negative infinite $q^2$ turns 
out to be the infinite volume limit. 

According to the original paper~\cite{Luscher:1985dn}, 
for {\it negative} $q^2$, 
we introduce the phase $\sigma_0(\kappa)$, which is defined
by an analytic continuation of $\delta_0$ into the complex $p$ plane 
through the relation $\tan \sigma_0(\kappa)=-i\tan \delta_0(p)$,
where $\kappa=-ip$. As a result, the bound-state pole condition 
in the infinite volume reads 
%
%
$\cot \sigma_0(\gamma)= -1$
%
for the binding momentum $\gamma$~\cite{Luscher:1985dn}.
Then, Eq.~(\ref{Eq:BScond}) can be rewritten in terms of the phase $\sigma_0$ as 
\be
\lim_{\kappa \rightarrow \gamma}
\cot \sigma_0(\kappa) 
= -1 +\sum_{\nu=1}^{\infty}\frac{N_{\nu}}{\sqrt{\nu} L\gamma}e^{-\sqrt{\nu}L\gamma} \\
= -1 + 
\frac{6}{L\gamma}\left[e^{-L\gamma}+{\cal O}(e^{-\sqrt{2}L\gamma})\right],
\label{Eq:BScondExp}
\ee
where the factor $N_{\nu}$ is the number of integer vectors ${\bf n}\in Z^3$
with $\nu={\bf n}^2$. Therefore, it is found that although a bound-state
pole condition is fulfilled only in the infinite volume limit, its modification 
by finite size corrections is exponentially suppressed by the spatial extent $L$
in a finite box $L^3$~\footnote[2]{Although it was pointed out how the bound-state pole
condition could be implemented in his phase-shift formula in the original 
paper~\cite{Luscher:1985dn}, this important fact has been 
firstly reported in Ref.~\cite{Sasaki:2006jn}.}.
We can learn from Eq.~(\ref{Eq:BScondExp}) that ``shallow bound states" 
are supposed to receive larger finite volume corrections
than those of ``tightly bound states" since the expansion parameter 
is scaled by the binding momentum $\gamma$. 


\section{Novel view from Levinson's theorem}

If the $S$-wave scattering length $a_0$, which is defined through 
$a_0=\lim_{p\rightarrow 0}\tan \delta_0(p)/p$, is sufficiently smaller than the spatial 
size $L$, one can make a Taylor expansion of the phase-shift formula
(\ref{Eq.LucsherFormula}) around $q^2=0$, and then obtain 
the asymptotic solution of Eq.~(\ref{Eq.LucsherFormula}).
Under the condition $p^2 \ll \mu^2$ where $\mu$ represents the reduced mass of 
two particles, the solution is given  by 
%
%
\be
\Delta E
\approx-\frac{2\pi a_0}{\mu L^3}\left[
1+c_1 \frac{a_0}{L}+c_2 \left(\frac{a_0}{L}\right)^2
\right] +{\cal O}(L^{-6}),
\label{Eq.ScattL0}
\ee
which corresponds to the energy shift of the lowest scattering state
from the threshold energy. The coefficients are $c_1=-2.837297$ and  
$c_2=6.375183$~\cite{Luscher:1985dn}.
An important message is received from Eq.~(\ref{Eq.ScattL0}).
The lowest energy level of the elastic scattering state appears below threshold  
($\Delta E<0$) on the lattice if an interaction is weakly attractive ($a_0>0$) 
between two particles. This point makes it difficult to distinguish between
near-threshold bound states and scattering states on the lattice.

Here, a crucial question arises: once the 
$S$-wave bound states are formed, what is the fate of the lowest $S$-wave 
scattering state? The answer to this question might provide 
a hint to resolve our main issue of how to distinguish between ``shallow bound states" 
and scattering states. A naive expectation from Levinson's theorem in quantum 
mechanics is that the energy shift relative to a threshold turns out to be opposite in 
comparison to the case where there is no bound state. 
Levinson's theorem relates the elastic scattering phase shift $\delta_l$ for 
the $l$-th partial wave at zero relative momentum to the total number of bound states ($N_{l}$) in a beautiful relation 
$\delta_{l}(0)=N_{l} \pi$~\footnote[2]{
Strictly speaking, this form is only valid unless zero-energy resonances exist.}.
Therefore, if an $S$-wave bound state is formed in a given channel, the 
$S$-wave scattering phase shift should always be positive at low energies. 
This positiveness of the scattering phase shift is consistent with a 
consequence of  the attractive interaction. 
Conversely, the $S$-wave scattering length may become negative ($a_0<0$), especially
for the shallow bound-states~\cite{Newton:1982qc}.
Consequently, according to Eq.~(\ref{Eq.ScattL0}), possible negativeness of the scattering length
gives rise to a positive energy-shift of the lowest scattering state relative to 
the threshold energy. In other words, the lowest scattering state
is pulled up into the region {\it above threshold}. 
Therefore, the spectra of the scattering states quite
{\it resembles the one in the case of the repulsive interaction}. 
If it were true, we can observe a significant difference 
in spectra above the threshold between the two systems: one has at least one 
bound state (bound system) and the other has no bound state (unbound system).

\section{Numerical results}

To explore signatures of bound-state formation on the lattice,
we consider a bound state (positronium) between an electron
and a positron in the compact QED with scalar matter~\cite{Sasaki:2006jn}:
%
%
\begin{equation}
S_{\rm SQED}[U,\Phi,\Psi]=
\beta \sum_{\rm plaq.}\left[1- \Re \{ U_{x, \mu \nu}\}\right]
-h\sum_{\rm link} \Re \{\Phi^{\ast}_x U_{x, \mu}
\Phi_{x+{\mu}}
\}
+\sum_{\rm sites}\overline{\Psi}_x D_{\rm W}[U]_{x,y} \Psi_y ,
\label{Eq:SQED}
\end{equation}
where $\beta=1/e^2$ and the constraint $|\Phi_{x}|=1$ is imposed.
This action is described by the compact $U(1)$ gauge theory coupled to 
both scalar matter (Higgs) fields $\Phi$ and fermion (electron) fields $\Psi$.  
In this study, we treat the fermion fields in the quenched approximation.

Our purpose is to study the $S$-wave bound state and
scattering states through L\"uscher's finite size method,
which is only applied to the short-ranged interaction case.
Thus, we fix $\beta=2.0$ and $h=0.6$ for the compact $U(1)$-Higgs action
to simulate the Higgs phase of $U(1)$ gauge dynamics, where massive photons 
give rise to the short-ranged interparticle force between an electron and a positron.
We generate $U(1)$ gauge configurations with a parameter set,
$(\beta, h)=(2.0, 0.6)$, on $L^3\times 32$ lattices
with several spatial sizes, $L=12,16, 20, 24, 28$ and 32.
Details of our simulations are found in Ref. ~\cite{Sasaki:2006jn}.

Once the parameters of  the compact $U(1)$-Higgs part, $(\beta, h)$, 
are fixed, the strength of an interparticle force between electrons 
should be frozen on given gauge configurations. However, if we 
consider the fictitious $Q$-charged electron, the interparticle force can be
controlled by this charge $Q$ since the interparticle force is 
proportional to (charge $Q)^2$. Within the quenched approximation, this trick 
of the $Q$-charged electron is
easily implemented by replacing  $U(1)$ link fields as
$U_{x,\mu} \longrightarrow U^{Q}_{x,\mu}=\Pi_{i=1}^{Q} U_{x, \mu}$
into the Wilson-Dirac matrix~\cite{Sasaki:2006jn}.
According to our previous pilot study~\cite{Sasaki:2005pc}, 
numerical simulations are performed with two parameter sets for fermion (electron) fields, $(Q, \kappa)$=(3, 0.1639) and (4, 0.2222). As we will see later, the former case ($Q=3$) corresponds to {\it the unbound system}, while the latter case ($Q=4$) 
corresponds to {\it the bound system} where the positronium state can be formed. 
Here $\kappa$, which is the hopping parameter
of the Wilson-Dirac matrix, is adjusted to yield almost the same 
electron masses $M_{e}\approx 0.5$ for both charges.

\begin{figure}
\begin{minipage}[t]{0.47\linewidth}
\bc
\includegraphics[angle=-90,width=0.9\textwidth]{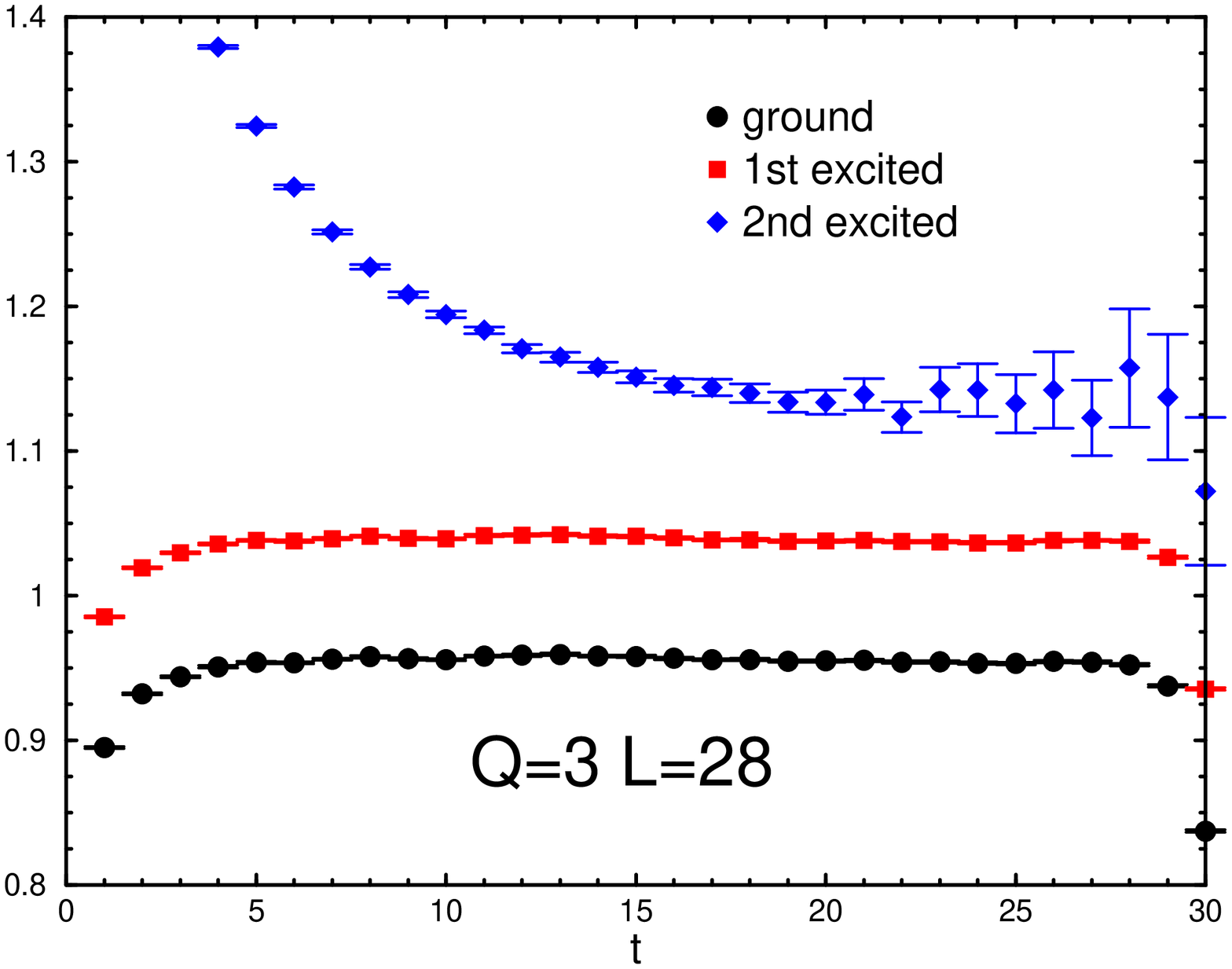}
\ec
\end{minipage}
 \hspace*{0.05\linewidth}
\begin{minipage}[t]{0.47\linewidth}
\bc
\includegraphics[angle=-90,width=0.9\textwidth]{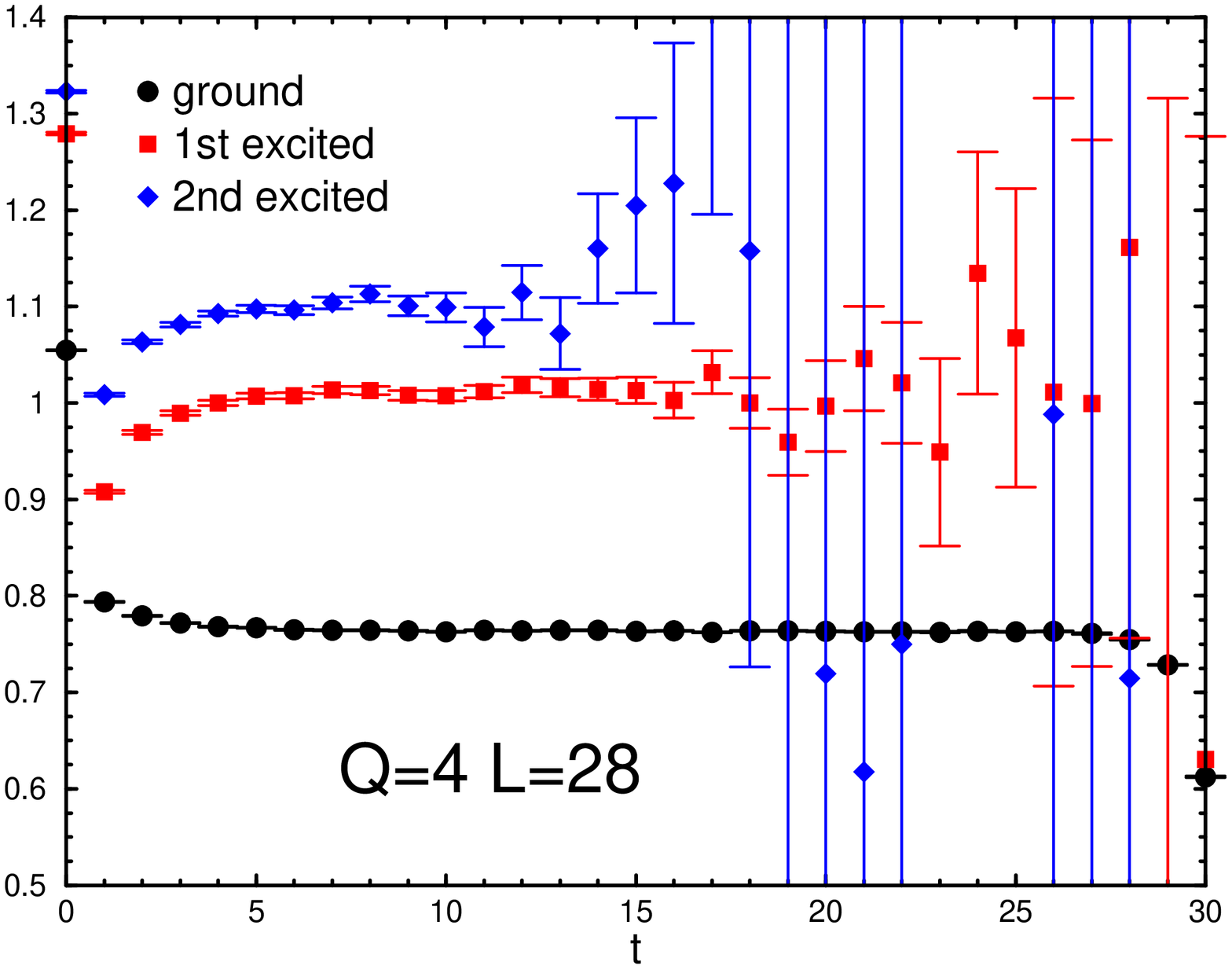}
\ec
\end{minipage}
\caption{The effective mass plots for each eigenvalue of the transfer matrix
in the $^1S_0$ channel on the lattice with $L=28$. Full circles, squares and diamonds represent  
the ground state, the first excited state and the second excited state. The left (right) panel is for $Q=3$ ($Q=4$).}
\label{FIG:Diag}
\end{figure}

We are especially interested in the $^1S_0$ and $^3S_1$ states of the $e^{-}e^{+}$ system,
where the electron-positron bound state (positronium) could be formed even in the Higgs phase.
$^1S_0$ and $^3S_1$ positronium are described by
the bilinear pseudo-scalar operator $\overline{\Psi}_x\gamma_5 \Psi_x$ and vector 
operator $\overline{\Psi}_x\gamma_\mu \Psi_x$ respectively. 
Therefore, we may construct the four-point functions of electron-positron states 
based on the above operators.
We are interested in not only the lowest level of two-particle spectra,
but also the 2nd and 3rd lowest levels. 
In order to extract a few low-lying energy levels of two-particle system, 
we utilize the diagonalization method~\cite{Luscher:1990ck}.
We consider three types of operators for this purpose:
$\Omega_P(t)={L^{-3}}\sum_{\bf x}{\overline \Psi}({\bf x}, t)\Gamma \Psi({\bf x},t)$,
$\Omega_W(t)={L^{-6}}\sum_{{\bf x},{\bf y}}{\overline \Psi}({\bf y}, t)\Gamma \Psi({\bf x},t)$
and
$\Omega_M(t)={L^{-6}}\sum_{{\bf x},{\bf y}}{\overline \Psi}({\bf y}, t)\Gamma \Psi({\bf x},t)e^{i{\bf p}_1\cdot({\bf x}-{\bf y}) }$ where ${\bf p}_1=\frac{2\pi}{L}(1,0,0)$ and 
$\Gamma=\gamma_5$  ($\gamma_{\mu}$) for the $^1S_0$  ($^3S_1$) $e^- e^+$ state. 
We construct the $3\times 3$ matrix correlator from above three operators
$G_{ij}(t)=\langle 0|\Omega_{i}(t)\Omega^{\dagger}_j(0)|0\rangle$
and then employ a diagonalization of a transfer matrix. As shown in Fig.~\ref{FIG:Diag}, 
the diagonalization method with our chosen three operators successfully 
separates the first excited state and the second excited state from 
the ground state~\cite{Sasaki:2006jn}.

%
%

\subsection{Sign of energy shift}
\begin{figure}
\begin{minipage}[t]{0.32\linewidth}
\bc
\includegraphics[width=0.90\textwidth]{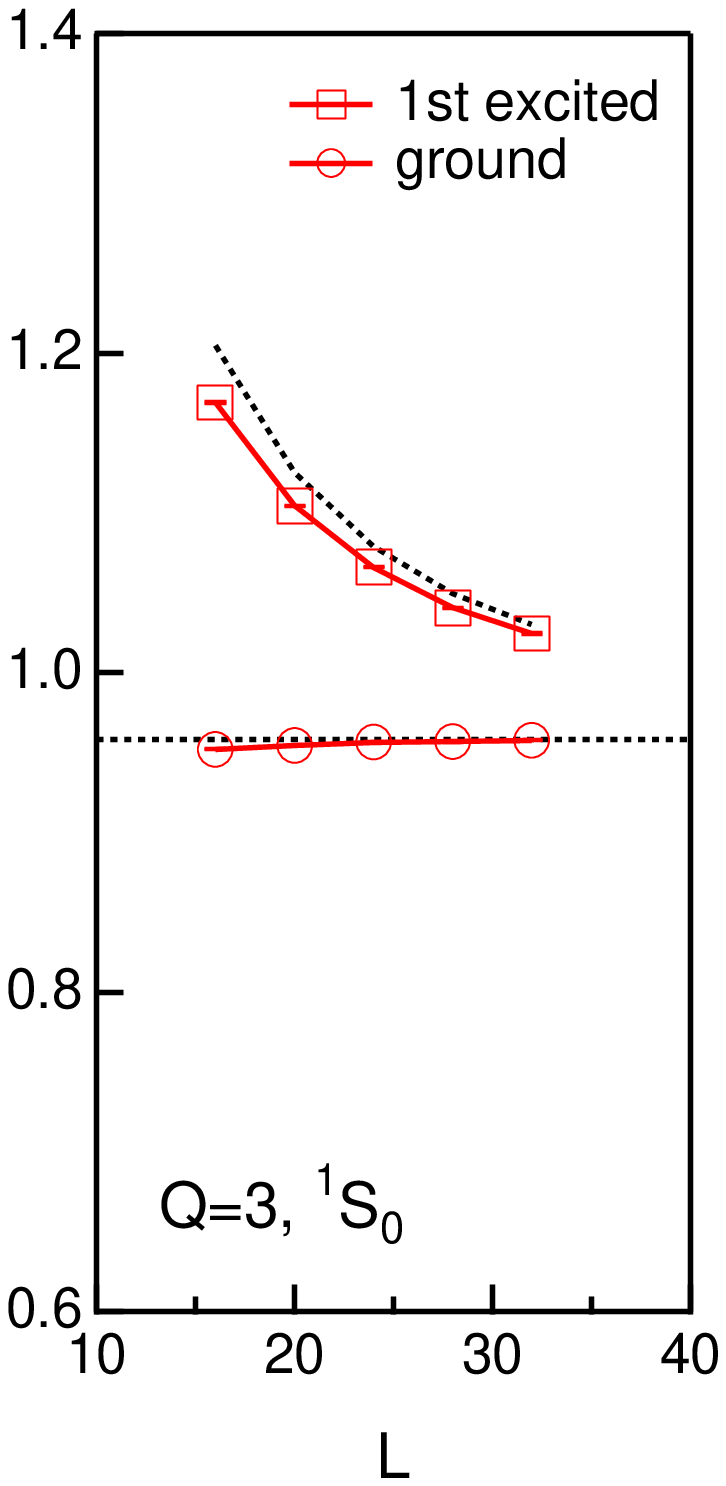}
\ec
\end{minipage}
\begin{minipage}[t]{0.32\linewidth}
\bc
\includegraphics[width=0.90\textwidth]{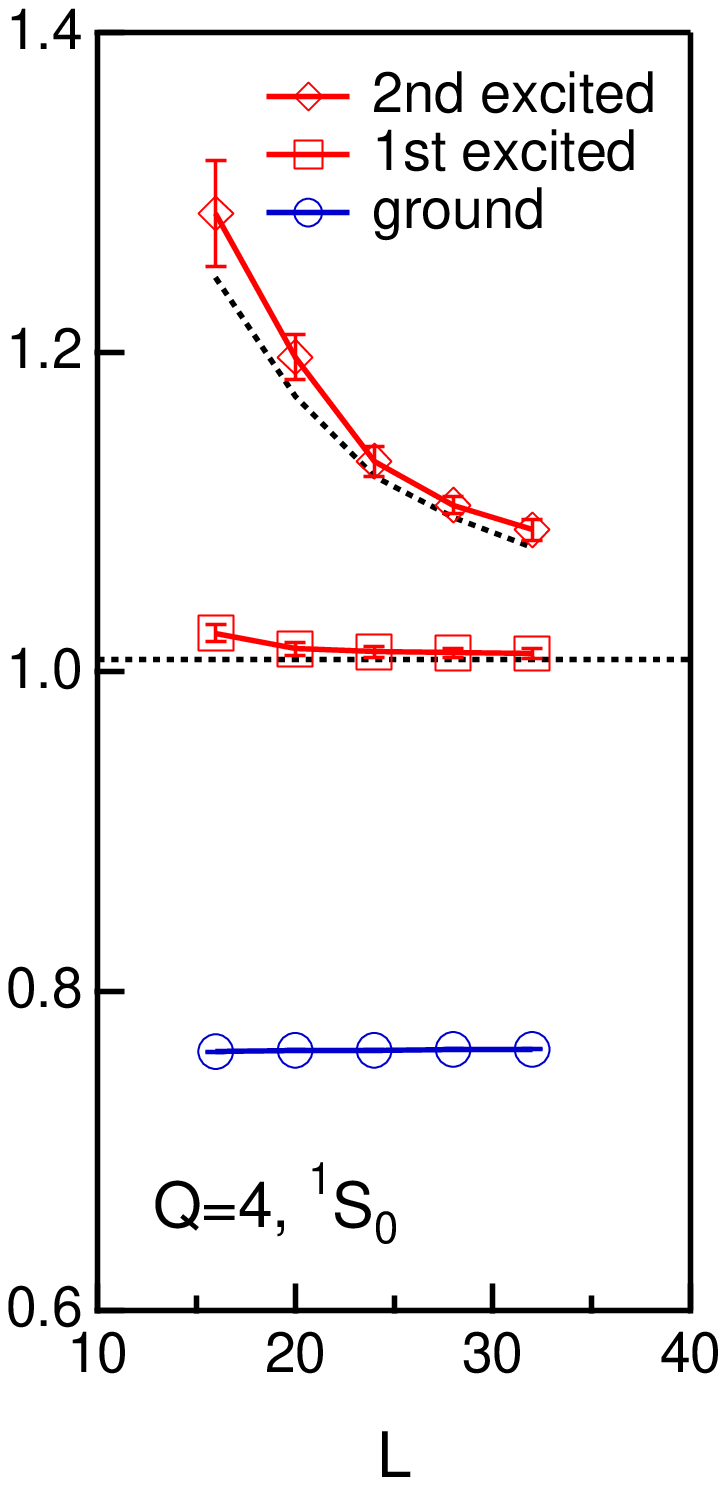}
\ec
\end{minipage}
\begin{minipage}[t]{0.32\linewidth}
\bc
\includegraphics[width=0.90\textwidth]{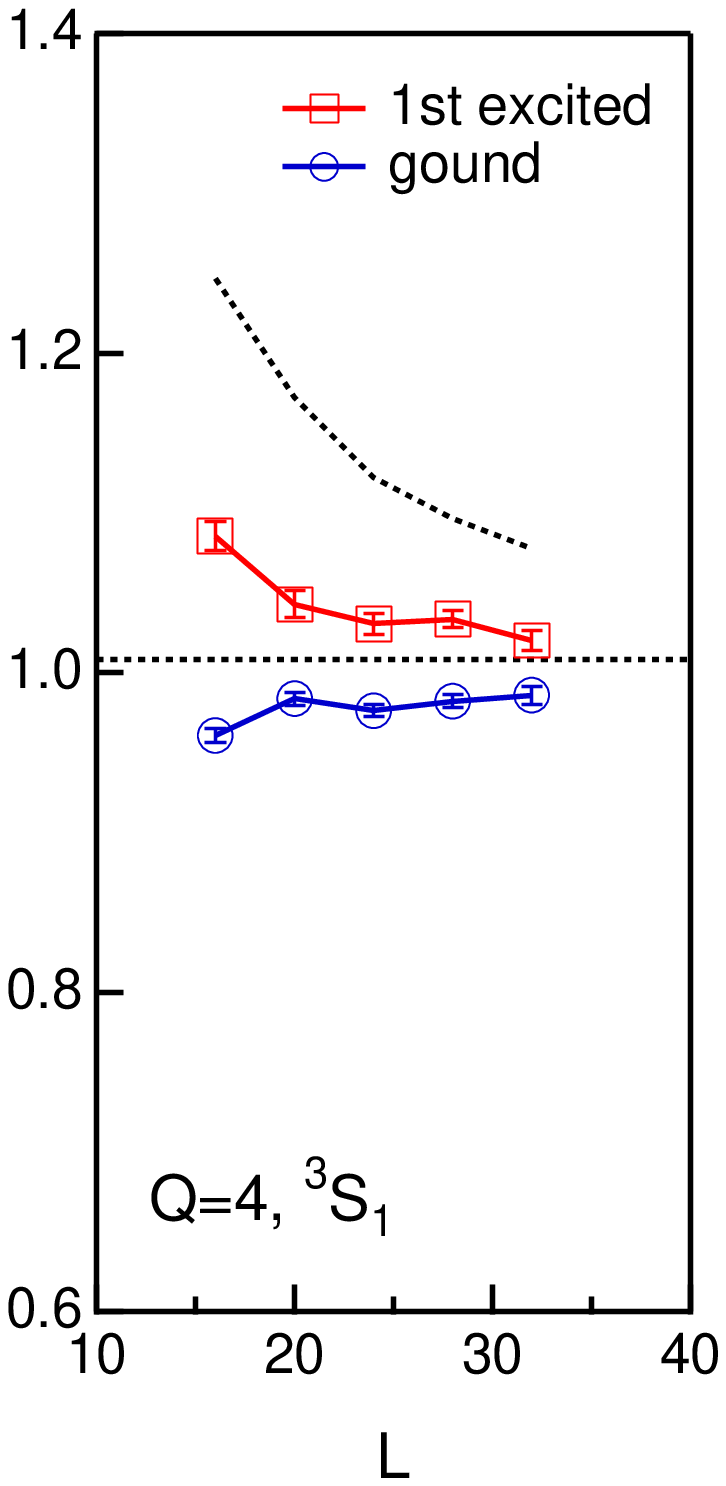}
\ec
\end{minipage}
\caption{Energies of the ground state and excited states 
in the $^1S_0$ (the left and middle panels) and $^3S_1$ (the right panel) channels 
of the $e^- e^+$ system as functions of spatial lattice size.
The left figure is for $Q=3$, while the middle and right panels are for $Q=4$. 
}
\label{FIG:Eflow}
\end{figure}

In Figs.~\ref{FIG:Eflow}, we show energies of the ground state and also excited states
in the $e^-e^+$ system as a function of spatial lattice size $L$. The dashed lines and
curves represent the threshold energies $2M_e$ and $2E_e({\bf p}_1)$, which are
evaluated by measured energies of the single electron with zero momentum ${\bf p}_0=\frac{2\pi}{L}(0,0,0)$
and  nonzero lowest momenta ${\bf p}_1=\frac{2\pi}{L}(1,0,0)$
respectively. Two left panels are for the $^1S_0$ channels, 
while the right panel is for the $^3S_1$ channel.

Let us focus on results in the $^1S_0$ channel. 
The energy level of the ground state for $Q=3$ in the left panel appears close
to the threshold. An upward tendency of the $L$-dependence toward the threshold
energy is observed as spatial size $L$ increases. This is consistent with a behavior
of the lowest scattering state predicted by Eq.~(\ref{Eq.ScattL0})
for the weakly attractive interaction without bound states. 
On the other hand, we clearly see the presence of a bound state 
for $Q=4$, which certainly 
remains finite energy gap from the threshold even in the infinite-volume limit.
The most striking feature is our observed $L$-dependence of the energy level 
of the second lowest state for ($Q=4$). Clearly, this energy
level approach the threshold energy {\it from above}. The energy shift vanishes
as the spatial size $L$ increases. Therefore, the second lowest energy state 
must be the lowest scattering state with the {\it repulsivelike}
scattering length ($a_0<0$). In addition, the level of the second lowest scattering state 
are located near and below (above) the threshold energy 
$2E_e({\bf p}_1)$ for $Q=3$ ($Q=4$).

In the $^1S_0$ channel for $Q=4$, the binding energy $B$ is rather large as 
$B\approx M_e/2$. The observed bound state should be a ``tightly bound state"
rather than a ``shallow bound  state". On the other hand, we observe that the bound
state in the $^3S_1$ channel (the right panel) is much near the threshold energy.
Although the $^3S_1$ ground state lies too close to the threshold energy to
be assured of bound-state formation, the distinctive signature of bound state
is given by an information of the excited state spectra. The second lowest state
appears just above the first threshold $2M_e$, but far from the second threshold
$2E_e({\bf p}_1)$. Therefore, we can conclude: the $^3S_1$ ground state
should be the shallow bound state, of which formation clearly induces
the sign of the scattering length to change~\cite{Sasaki:2006jn}.

%
%

\subsection{Bound-state pole condition}

\begin{figure}
\begin{minipage}[t]{0.47\linewidth}
\bc
\includegraphics[angle=-90,width=0.9\textwidth]{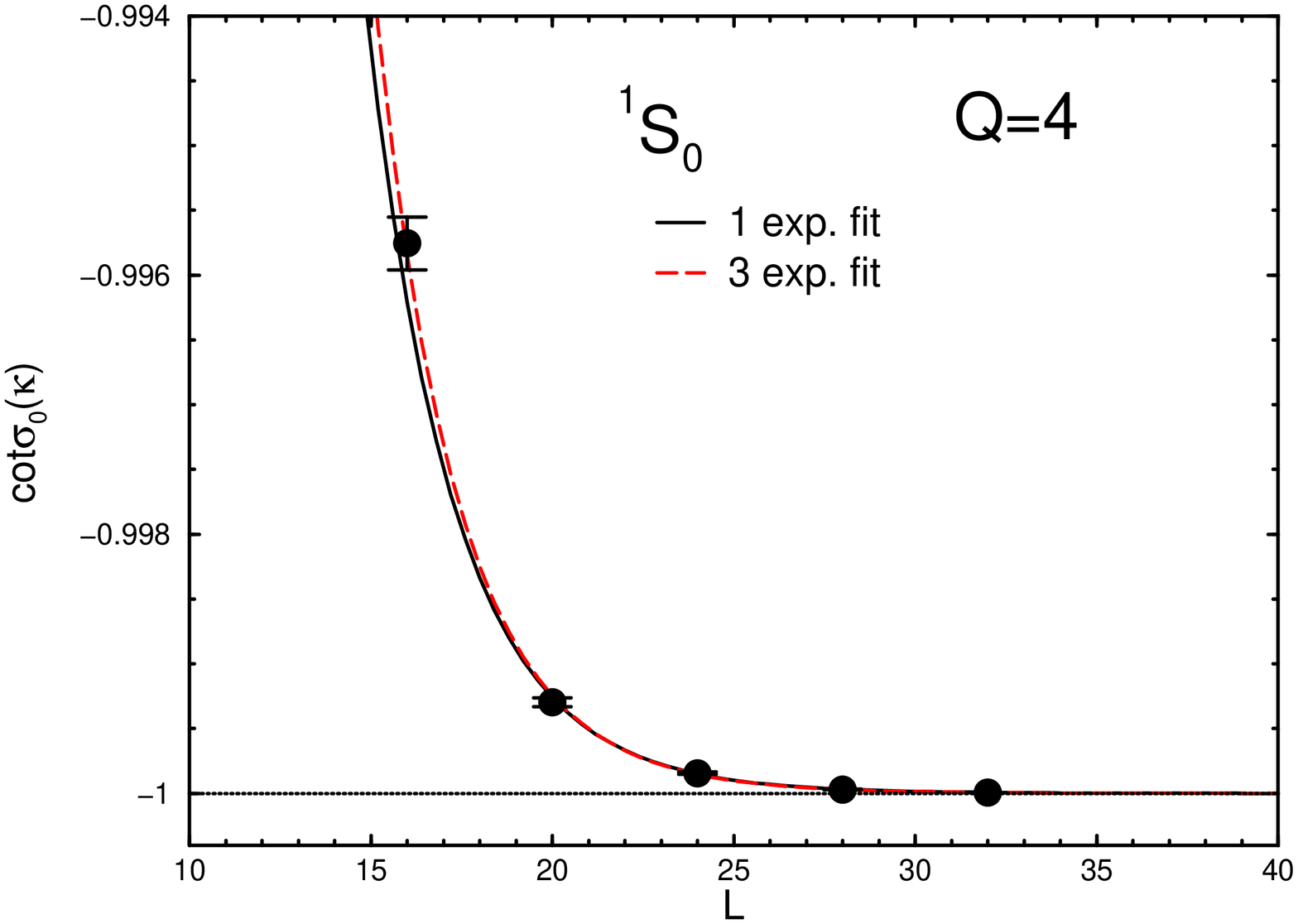}
\ec
\end{minipage}
 \hspace*{0.05\linewidth}
\begin{minipage}[t]{0.47\linewidth}
\bc
\includegraphics[angle=-90,width=0.9\textwidth]{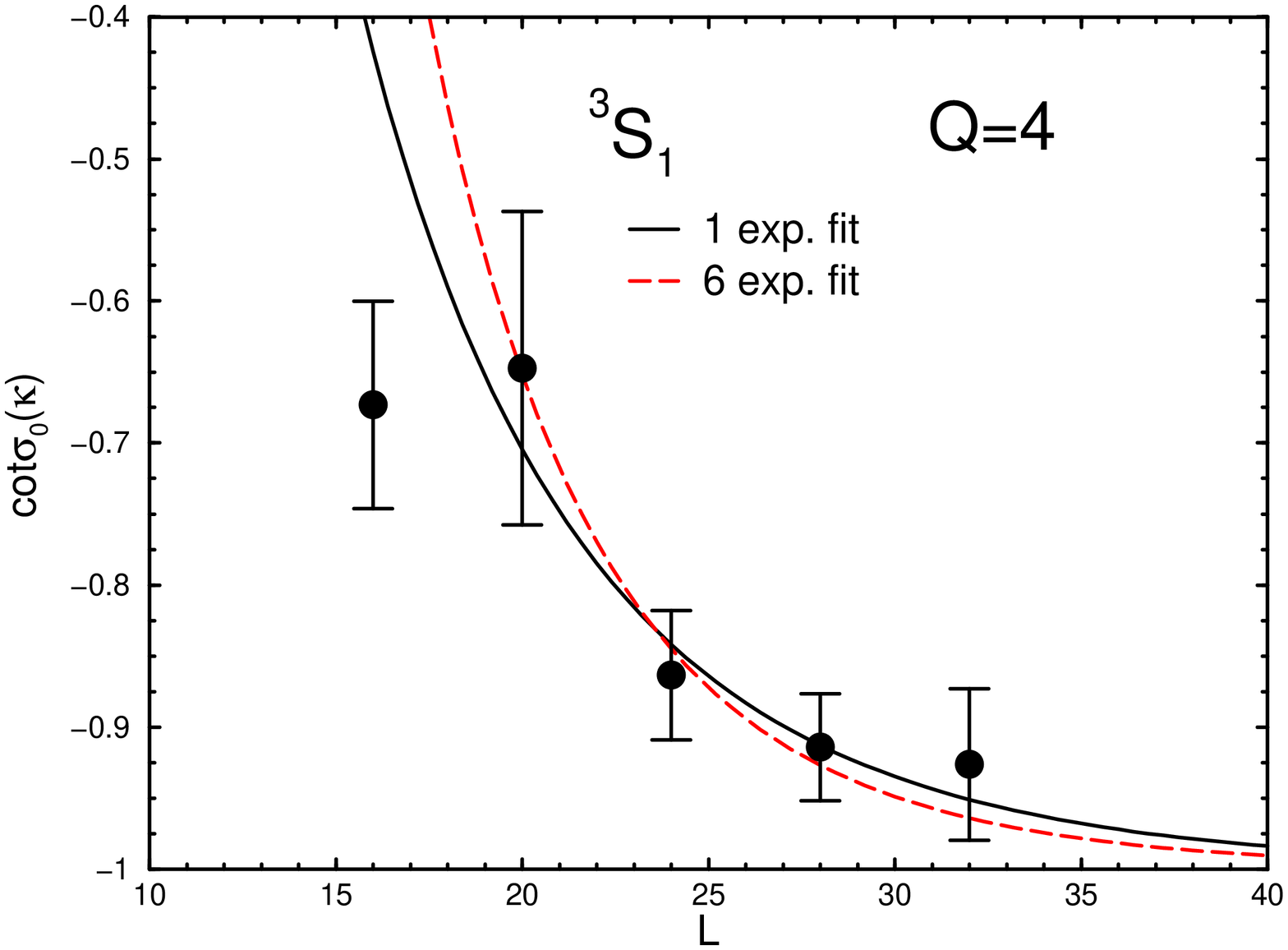}
\ec
\end{minipage}
\caption{$\cot \sigma_0$ in the $^1S_0$ (left) and $^3S_1$ (right) channel for
$Q=4$ as a function of the spatial lattice size $L$.}
\label{FIG:cot}
\end{figure}

A rigorous way to test for bound-state formation would be to use an asymptotic formula
for finite volume correction to the pole condition as Eq.~(\ref{Eq:BScondExp}). 
In Figs.\ref{FIG:cot}, we plot the value of $\cot \sigma_0$ versus the spatial lattice extent $L$ 
for either $^1S_0$ (left) and $^3S_1$ (right) channels. Full circles are measured
value at five different lattice volumes. At first glance, we observe 
that the phase $\cot \sigma_0$ gradually approaches $-1$ as spatial lattice extent $L$ increases for either channels. 

We next examine the $L$-dependence of $\cot \sigma_0$ by reference to  Eq.~(\ref{Eq:BScondExp}), where the finite volume corrections on the bound-state pole condition are theoretically predicted. The solid and dashed curves represent fit results with a single 
leading exponential term and three (six) exponential terms in the $^1S_0$ ($^3S_1$)
channel. All five data points are used for those fits in the $^1S_0$ channel, while
the four data points in the region $20\le L \le 32$ are used in the $^3S_1$ channel.
The fitting with the three (six) exponential terms yields a convergent result of $\gamma$
in the $^1S_0$ ($^3S_1$) channel. Either fit curves in Figs.~\ref{FIG:cot}
reproduce all data points except for data at the smallest $L$ in the $^3S_1$ channel.
Therefore, we confirm that the ground state in the $^3S_1$ channel
at least for $L\ge 20$ can be identified as a shallow bound state without ambiguity.

%
%

\section{Summary and conclusion}

In this paper, we have discussed 
formation of an S-wave bound-state in finite volume on the basis
of L\"uscher's phase-shift formula. We have first showed that although a bound-state
pole condition is fulfilled only in the infinite volume limit, its
modification by the finite size corrections is exponentially suppressed by
the spatial extent $L$ in a finite box $L^3$. 
We have also confirmed that the appearance of the S-wave bound state is accompanied 
by an abrupt sign change of the S-wave scattering length even in finite volume through 
numerical simulations. This distinctive behavior may help us to discriminate the
shallow bound state from the lowest energy level of the scattering state in
finite volume simulations.



\begin{thebibliography}{99}

  
\bibitem{Luscher:1985dn}
  M.~L\"uscher,
  Commun.\ Math.\ Phys.\  {\bf 104}, 177 (1986),
  Nucl.\ Phys.\ B {\bf 354}, 531 (1991).

\bibitem{Newton:1982qc}
  R.~G.~Newton, 
  ``Scattering Theory of Waves and Particles'', 2nd ed. (Springer, New York, 1982).

\bibitem{Sasaki:2006jn}
  S.~Sasaki and T.~Yamazaki,
  Phys.\ Rev.\  D {\bf 74}, 114507 (2006).

\bibitem{Elizalde:1997jv}
  E.~Elizalde,
  Commun.\ Math.\ Phys.\  {\bf 198}, 83 (1998).
   
  
\bibitem{Sasaki:2005pc}
  S.~Sasaki and T.~Yamazaki,
  PoS {\bf LAT2005}, 061 (2006).

\bibitem{Luscher:1990ck}
  M.~L\"uscher and U.~Wolff,
  Nucl.\ Phys.\ B {\bf 339}, 222 (1990).
  
\end{thebibliography}
\end{document}